\documentclass[12pt]{article}

\usepackage{graphics}
\usepackage{graphicx}
\usepackage{subcaption} 
\usepackage{color}
\usepackage{amssymb}
\usepackage{amsmath}
\usepackage{verbatim}

\usepackage[colorlinks = true,
linkcolor = blue,
urlcolor  = blue,
citecolor = blue,
anchorcolor = blue]{hyperref}

\usepackage[numbers,square,sort&compress]{natbib}
\usepackage{authblk}
\usepackage{cancel}

\numberwithin{equation}{section}
\begin{document}
	\title{\bf Constraints on anisotropic RG flows from holographic entanglement entropy} 
	\author[1]{Mostafa Ghasemi}
	\author[2]{Shahrokh Parvizi}
	\affil[1]{ School of Particles and Accelerators, Institute for Research in Fundamental Sciences (IPM)\\
		P.O. Box 19395-5531, Tehran, Iran}
	\affil[2]{ Department of Physics, School of Sciences,
			 Tarbiat Modares University, P.O.Box 14155-4838, Tehran, Iran} 
	\affil[ ]{\textit {Email: \href{mailto:ghasemi.mg@ipm.ir}{ghasemi.mg@ipm.ir},   \href{mailto:parvizi@modares.ac.ir}{parvizi@modares.ac.ir}}}
	\date{\today}                     
	\setcounter{Maxaffil}{0}
	\renewcommand\Affilfont{\itshape\small}
	
	\maketitle
	
	\abstract 
	
In the context of the gauge/gravity duality, using the proposed candidate $c$-function, which is derived from the entanglement entropy of a strip-shaped region, we investigate the RG flow for $d+1$-dimensional quantum field theories with broken Lorentz and rotational symmetries in the IR, but preserved conformal invariance in the UV boundary. We examine conditions of monotonicity of the $c$-function for holographic anisotropic theories dual to the Einstein gravity via the constraints imposed by the null energy conditions. We consider near UV and IR behaviors and identify the sufficient conditions that guarantee the $c$ function decreases monotonically along the RG flows.

	
	\vspace{10mm}
	
	\newpage
	\section{Introduction} \label{intro}
	A challenging question in the quantum field theory (QFT) is finding a proper quantity in order to probe the structure of parameter space of QFT. The renormalization group (RG) provides a framework to study various aspects of the QFT space. The RG denotes how physics in different energy scales can be related to each others.
		 
	There are special points in this space, called fixed points, which  correspond to theories with scaling symmetry. Other theories may be derived from such points by various deformations. By definition, this quantity/function is definite positive and monotonically decreases along the RG flows and is a measure of degrees of freedom of the effective theory. Another feature of this function is that its values at fixed points equal to the central charges of the corresponding theories.
		
    In the space of two-dimensional QFT, Zamolodchikov \cite{Zamolodchikov:1986gt} showed that there exists such a positive definite function $c_{2}$, which decreases monotonically along the RG flows.
	This function is stationary and coincides with the central charge $c$ of the
	associated fixed points CFT. As a consequence, if RG flows connect two UV and IR fixed points, we have
		\begin{equation}
		  c_{UV}>c_{IR}.
		\end{equation}
	From the Wilsonian point of view, it is a measure of the effective degrees of freedom that decreases along the RG flows. 
	
	One of the suitable candidates $c$-function is emergent in the realm of the information theory. In this context, the proper $c$-function is derived from the entanglement entropy that is a measure of quantum entanglement. It can be used as a measure of effective degrees of freedom in QFT along the RG flows in the Wilsonian sense. For a two-dimensional conformal field theory, the entanglement entropy for an interval of length $l$ is given by \cite{Calabrese:2004eu,Calabrese:2009qy}
	\begin{equation}
		  S_{EE}^{CFT}=\frac{c}{3}\log (\frac{l}{\delta})+\cdots,
	\end{equation}
	where $c$ is the central charge, $\delta$ is a UV regulator, and dots denote the $l$-independent non-universal terms.
		
	The authors \cite{Casini:2004bw,Casini:2006es} reformulated the
	Zamolodchikov's c-theorem for a two-dimensional QFT in terms of $S_{EE}$
	and defined	
	\begin{equation}
		  c_{2}=3l\frac{dS_{EE}(l)}{dl}.
	\end{equation}
	The monotonicity flow of the $c_{2}$, $dc_{2}/dl\leq0 $, can be shown from the strong subadditivity property of $S_{EE}$ as well as the Lorentz symmetry and unitarity of the underlying QFT. It takes the value $c_{2}=c$ at RG fixed points.
		
The generalization of the $c$-theorem for higher dimensions is remained a challenging problem. This issue has been addressed in \cite{Cardy:1988cwa,Komargodski:2011vj,Komargodski:2011xv,Casini:2012ei} for three- and four-dimensional theories with Lorentz invariance. On the other hand, the gauge/gravity duality provides a framework to investigate the RG flows to arbitrary dimensions in the gravitational context \cite{Freedman:1999gp,Girardello:1998pd,Girardello:1999bd,Myers:2010xs,Myers:2010tj}. The connection between RG flows and the entanglement entropy well is depicted via holographic prescription
of computing the entanglement entropy of the boundary theory \cite{Ryu:2006bv,Ryu:2006ef,Nishioka:2018khk}.  The authors \cite{Ryu:2006ef,Myers:2012ed} defined a candidate $c$-function which extracted from the holographic entanglement entropy of a strip-shape region. In \cite{Myers:2012ed}, it was shown that the $c$-functions for a Lorentz invariant QFT decreases monotonically if the bulk gravity/matter fields satisfy the null-energy conditions.
		
There is a natural question that in what extend one can recover monotonic RG flows for theories which exhibit broken Lorentz invariance.  In \cite {Swingle:2013zla}, it was shown that for weakly coupled Lorentz-violating field theories, the entanglement entropy does not decrease monotonically under RG flows. On the other hand, \cite {Cremonini:2013ipa} claimed that for Lorentz-violating strongly coupled field theories with holographic dual such a breakdown of the candidate $c$ function can be seen. However they also, with respect to the asymptotic UV behavior of the geometry, identified sufficient conditions to have a monotonically decreasing $c$ function along the RG flow. These theories are associated to geometries with a dynamical exponent $z$ and a hyperscaling violation parameter $\theta$.

These dual models are important in studying QFTs with violation of Lorentz invariance and broken rotational symmetry \cite{Giataganas:2017koz,Giataganas:2018ekx,Giataganas:2018rbq,Jain:2015txa,Giataganas:2013lga,Giataganas:2012zy,Gursoy:2018ydr,Gubser:2008ny,Gubser:2008yx,Ghasemi:2019hdi,Alishahiha:2012cm}. It is worth mentioning that there are studies on RG-flows in theories without Lorentz symmetry while they have Galilean invariance only. There is a possibility to introduce monotonically decreasing quatities in these models which start from a UV fixed point which is not relativistically conformal invariant, so they are anisotropic models with Lifshitz symmetry 
\cite{Pal:2016rpz,Arav:2017plg,Auzzi:2016lrq,Solodukhin:2013yha}.
		
In this work, we raise the question that how the anisotropy affects the monotonicity of the RG flows, and in what conditions we can define the candidate c-function from holographic entanglement entropy for theories that exhibit Lorentz-violating as well as rotational symmetry breaking. We will construct a $c$ function that interpolates between a UV CFT and an IR anisotropic theory.

For this regarding, we consider the cases  in which the Lorentz and rotational symmetries are broken only in the IR, while the UV boundary is still conformally invariant, and then examine flowing from a fixed point and, in particular, a c-function of the RG flow.
There are other related works in this context \cite{Ghosh:2018qtg, Liu:2019npm}. 
		
This paper is organized as follows. In section \ref{sec:EE}, first we briefly review the holographic entanglement entropy of a strip-shaped region in an $AdS$ background, then we derive the Ryu-Takayanagi minimal surface for an anisotropic dilaton-axion model and also in a hyperscaling violation background. In section \ref{sec:c-function}, we derive null energy conditions, and propose the candidate $c$-function, for geometries dual to anisotropic QFTs. In section \ref{sec:examples}, we illustrate our results in some anisotropic backgrounds. Finally we conclude in section \ref{sec:conclusion}.
In appendix \ref{app:EE}, we provide more details on calculating the holographic entanglement entropy. To do so, we take the width of the strip in two directions: first, along the isotropic scaling dimensions, then in the anisotropic direction. More details on deriving the RG-flow are provided in appendix \ref{app1}.

	{\bf Note added:}
		While this article was being completed, we received \cite{Chu:2019uoh} which has some overlaps with our results.
	
\section{Holographic entanglement entropy of anisotropic models}\label{sec:EE}
In the following subsections, we calculate the holographic entanglement entropy of a strip-shaped region for some anisotropic models. To understand its relationship to the c function, let us first recall \cite{Myers:2012ed} approach, which leads to the identification of the candidate c-function by the holographic entanglement entropy of a strip-shaped region for a CFT. 

In the UV fixed point we have a CFT, which is dual to $AdS_{d+2}$ background. Therefore, we can apply Ryu-Takayanagi ($RT$) prescription \cite{Ryu:2006bv,Ryu:2006ef} to derive the holographic entanglement entropy.
For a sub-region $V$ on the $d+1$-dimensional boundary field theory, it is given by
\begin{equation}
  S_{EE}= \frac{Area(m)}{4G_{N}},
\end{equation}
where $m$ is the minimal surface in $d+2$-dimensional bulk and is homologous to $V$ and $\partial_{m}$ matches the entangling surface $\partial_{V}$ on the boundary. The above formula holds in the case that the bulk physics is described by the Einstein gravity.

The holographic entanglement entropy for a strip-shaped region, where entangling surface is described by two parallel $(d-1)$-dimensional planes separated by a distance $\ell$, is given by \cite{Ryu:2006bv,Ryu:2006ef}
\begin{equation}\label{Scft}
S_{CFT}=\alpha_{d}\Big(\frac{H}{\delta}\Big)^{d-1}-\frac{1}{(d-1)\beta_{d}} C_{d}\Big(\frac{H}{\ell}\Big)^{d-1}
\end{equation}
where $\alpha_{d}$ and $\beta_{d}$ are positive dimensionless numerical factors, and $H\gg\ell$ can be intended as an IR regulator along the entangling surface. The first term denotes the area law and . the second one is finite with coefficient $C_{d}$, which is related to the central charge of the underlying CFT via
\begin{equation}
C_{d}=\beta_{d}\frac{\ell^{d}}{H^{d-1}}\frac{\partial S_{EE}^{CFT}}{\partial \ell}
\end{equation}
 
This is suggestive that a c-function candidate along the RG flows can be extracted from the holographic entanglement entropy as \cite{Ryu:2006ef,Myers:2012ed}
  \begin{equation}\label{cd}
\mathit{c}_{d}=\beta_{d}\frac{\ell^{d}}{H^{d-1}}\frac{\partial S_{EE}}{\partial \ell}.
 \end{equation}
where at the fixed point we have $\mathit{c}_{d}=\mathit{C}_{d}$. The monotonicity of this function along the RG flows comes from the Lorentz symmetry and subadditivity inequalities of the entanglement entropy. 

In the next section, we promote it to an anisotropic theory and consider the sufficient conditions for monotonically decreasing of the candidate c-function along RG flows. Before that, it would be helpful to find the entanglement entropy of a strip-shaped region for such theories. We consider two kinds of anisotropic models.

\subsection{An anisotropic model in the UV regime}\label{subsec:axion-dilaton}
In this subsection, we consider an anisotropic theory for which the dual gravitational theory is defined by the Einstein-axion-dilaton action  \cite{Gubser:2008ny,Gubser:2008yx,Giataganas:2017koz}, 
\begin{align}\label{axion-dilaton}
	S&=\frac{1}{2\ell_p^3}\int d^5x\sqrt{-g}\Big(R+\mathcal{L}_M\Big) \nonumber\\
	\mathcal{L}_M&=-\frac{1}{2}(\partial\phi)^2+V(\phi)-\frac{1}{2}Z(\phi)(\partial\chi)^2 
\end{align}
where $V$  and $Z$ are a potential for the dilaton field $\phi$, and a coupling between the axion field $\chi$ and the dilaton, respectively. With respect to generic choice of the $V$ and $Z$ we can find various solutions. With
\begin{align}\label{Full-potential}
	V(\phi)&=12\cosh(\sigma\phi)-6\sigma^2\phi^2, \qquad Z(\phi)=e^{2\gamma\phi} 
\end{align} 
near the boundary, this potential approaches the cosmological constant (with the radius of curvature is $L=1$) and we have a UV fixed point with conformal symmetry on the boundary.

Now take a linear axion as $\chi=a y$, then the Einstein's equations can be found as
\begin{align}
	3 A' \left(2 A'+h'\right)&=\frac{1}{4} \left(2 e^{2 A} V(\phi)-a^2e^{-2 h} Z(\phi)+\phi'^2\right)  \\
	3 A''+3 A' h'+3 A'^2+h''+h'^2& =\frac{1}{4} \left(2 e^{2 A} V(\phi)-a^2e^{-2 h} Z(\phi)-\phi'^2\right) \\
	3 e^{2 h} \left(A''+A'^2\right) &=\frac{1}{4} \left(2 e^{2 (A+h)} V(\phi)-e^{2 h} \phi'^2+a^2Z(\phi)\right)
\end{align}
and the dilaton equation,
\begin{align}
	\phi''+\phi' \left(3 A'+h'\right)=a^2 \gamma  e^{2 \gamma  \phi-2 h}+12 \sigma  e^{2 A} (\sigma  \phi-\sinh (\sigma  \phi)).
\end{align}
where an anisotropic ansatz is introduced as
\begin{align}
	ds^2&=e^{2A(r)}\Big[-f(r)dt^2+\frac{dr^2}{f(r)}+dx_1^2+dx_2^2+e^{2h(r)}dy^2\Big]\nonumber\\
	\phi&=\phi(r), \qquad \chi=a y.
\end{align}
Here, we restrict ourselves to zero temperature solution for which $f(r)=1$. 
For a small anisotropy $ar\ll 1$, a perturbative solution can be found as
\begin{align}\label{small-a}
	A(r)&=-\log(r) - \frac{a^2 r^2}{72}+\frac{a^4 r^4}{1200}\left(3 \gamma ^2+1\right)  (1-5 \log (ar))+\mathcal{O}(ar)^6, \\
	h(r)&=\frac{a^2r^2}{8}-\frac{a^4 r^4}{2592}\Big(31+81 \gamma ^2-54 \left(3 \gamma ^2+1\right) \log (ar)\Big)+\mathcal{O}(ar)^6,\\
	\phi(r)&=-\frac{\gamma a^2r^2}{4}+\frac{ a^4 r^4}{96} \gamma(3 \gamma ^2+1)(1-4 \log (ar)) +\mathcal{O}(ar)^6.
\end{align}

In the following, we are going to find the entanglement entropy for two cases with the width of strip along one of $x$ coordinates and along $y$ direction.

Let us first consider the width of the strip along one of $x_i$'s directions, say $x=x^1$ as $-\ell/2 \leq x\leq \ell/2$, then parametrize the RT surface as $r=r(x)$. The entanglement entropy follows as
\begin{align}\label{EE}
	S_x=\frac{4\pi H^{d-1}}{\ell_p^d}\int_0^{\frac{\ell-\epsilon}{2}}dx e^{B}\sqrt{\dot{r}^2+1}
\end{align} 
where $B(r)=dA(r)+h(r)$ with $d=3$, $\epsilon$ is a UV cutoff and $H$ is a cutoff on length of the strip.
The EoM is
\begin{align}\label{EoM}
	\ddot{r}=B'(1+\dot{r}^2)
\end{align}
where dot denotes $d/dx$ and prime is $d/dr$. Since the integrand of \eqref{EE}, $\mathcal{L} $ is independent of $x$, we can introduce the following constant of motion,
\begin{align}\label{K}
	K_d(\ell)&=\Big(\mathcal{L}-\dot{r}\frac{\partial\mathcal{L}}{\partial\dot{r}}\Big)^{-1}\nonumber\\
	&=e^{-B}\sqrt{\dot{r}^2+1}
\end{align}
Then we can write
\begin{align}
	K_d(r_m)=e^{-B(r_m)}
\end{align}
where $ r_m $ is the maximum of $r $ at $\dot{r}=0$.  
It follows then
\begin{align}\label{SEE}
	S_x&=\frac{4\pi H^{d-1}}{\ell_p^{d}}\int_0^{\frac{\ell-\epsilon}{2}}dx e^{2 B(r(x))-B(r_m)}   \nonumber\\
	&=\frac{4\pi H^{d-1}}{\ell_p^{d}}\int_\delta^{r_m}dr \frac{e^{2 B(r(x))-B(r_m)}}{\sqrt{e^{2(B(r)-B(r_m))}-1}}
\end{align}
where we used $\dot{r}=-\sqrt{e^{2 B(r)-2 B(r_m)}-1}$ and $\delta$ is the UV cutoff on $r$ coordinate. 

We compute the entanglement entropy in the appendix \ref{app:EE} and find it as,
\begin{align} \label{Sell2}
	S_x=\frac{4\pi H^{2}}{\ell_p^3} \Big[&-\frac{0.1603}{\ell^2}+\frac{a^2}{12}\log \ell
+0.0316 a^2-(0.0088 +0.0293 \gamma ^2+ 0.0048 \gamma ^4) a^4 \ell^2 \nonumber\\
&+(0.0079+0.0127 \gamma ^2-0.0326 \gamma ^4) a^4 \ell^2 \log (a\ell) \Big]. 
\end{align} 
The leading term corresponds to \eqref{Scft} with $d=3$, but in contrast, the second term in \eqref{Sell2} is a universal logarithmic term which has no counterpart in \eqref{Scft}. 


As the second case, we consider the strip as $-\ell/2 \leq y\leq \ell/2$, and reparametrize the RT surface as $r=r(y)$. Then we have
\begin{align}\label{EEy}
	S_y=\frac{4\pi H^{d-1}}{\ell_p^{d}}\int_0^{\frac{\ell-\epsilon}{2}}dx e^{dA}\sqrt{\dot{r}^2+e^{2h}}.
\end{align} 
The constant of motion is
\begin{align}\label{Ky}
	K_d(\ell)&=e^{-B}\sqrt{e^{-2h}\dot{r}^2+1},\qquad \implies \qquad K_d(r_m)=e^{-B(r_m)}
\end{align}  
It follows then
\begin{align}\label{SEEy}
	S_y&=\frac{4\pi H^{d-1}}{\ell_p^{d}}\int_0^{\frac{\ell-\epsilon}{2}}dy e^{2 B(r(y))-B(r_m)}   \nonumber\\
	&=\frac{4\pi H^{d-1}}{\ell_p^{d}}\int_\delta^{r_m}dr \frac{e^{2 B(r)-B(r_m)}}{e^{-h}\sqrt{e^{2B(r)-2B(r_m)}-1}}
\end{align}
where we used $\dot{r}=-e^{-h}\sqrt{e^{2 B(r)-2 B(r_m)}-1}$.
From the calculations in the appendix  \ref{app:EE}, we found,
\begin{align} \label{Sell2y}
	S_y=\frac{4\pi H^{2}}{\ell_p^3} \Big[&-\frac{0.1603}{\ell^2}-\frac{a^2}{24} \log (\ell)	
	+0.0259 a^2+(0.0002+0.0050 \gamma^2) a^4 \ell^2   \nonumber\\
	&-(0.0039+0.0118 \gamma ^2) a^4 \ell^2 \log (a\ell)\Big] 
\end{align}

\subsection{Anisotropy in a hyperscaling model}\label{subsec:hyper}

As another interesting case, consider the potential $V$ in \eqref{axion-dilaton} to be \cite{Giataganas:2017koz, Giataganas:2018ekx}
\begin{align}\label{exp-potential}
	V(\phi)&=6e^{\sigma\phi}, \qquad Z(\phi)=e^{2\gamma\phi} 
\end{align}
the resulting solution of equation of motions is a Lifshitz-like anisotropic hyperscaling violation metric with the arbitrary critical exponent $z$. The hyperscaling violation exponent $\theta$ is related to the $\sigma$ and $\gamma$ constants.  It is worth mentioning that this potential can be considered as the IR limit of \eqref{Full-potential}.
But, these solutions indeed exist as exact metrics along UV to IR (not only as limits). 
The Lagrangian setup for these solutions from UV to IR as well as  the exact scaling factors of the background solution are presented in \cite{Giataganas:2018ekx,Giataganas:2018rbq}\footnote{We would like to thank Dimitrios Giataganas for pointing this out to us.}. 
There are other related works in this background and context \cite{Gursoy:2018ydr, Ghasemi:2019hdi}.
In otherwise, the anisotropic IR metrics that is derived in \cite{Giataganas:2017koz}  can be understood as a double wick rotation of the standard Lifshitz/HsV metrics as well as scaling of $z$ and $\theta$ \cite{Alishahiha:2012cm}.

In any way, here we suppose it to be an independent potential valid everywhere (not only IR). Now by a linear axion in the $y$-direction, $\chi=a y$, one finds the following exact solution \cite{Giataganas:2017koz, Giataganas:2018ekx}
\begin{align}\label{metric0}
	ds ^{2}&= \tilde{L}^{2}(ar)^{\frac{2\theta}{dz}}(\frac{-dt^{2}+dr^{2}+d\vec{x}_{d-1}^{2}}{(ar)^{2}}+\frac{c_1dy^{2}}{(ar)^{2/z}}),\nonumber\\
	\phi&=c_2\log(ar)+\phi_0. 
\end{align}
where $z=(4 \gamma ^2-3 \sigma ^2+2)/(4 \gamma^2- 6\gamma \sigma )$ and $\theta=3 \sigma/(2 \gamma)$. For simplicity we consider $a=1$ and absorb $c_1$ in $y$ coordinate,  \begin{equation}\label{metric1}
	ds ^{2}= \tilde{L}^{2}r^{\frac{2\theta}{dz}}(\frac{-dt^{2}+dr^{2}+d\vec{x}_{d-1}^{2}}{r^{2}}+\frac{dy^{2}}{r^{2/z}}).
\end{equation}
The scale transformation of the coordinates is defined as 
\begin{equation}\label{scaling}
	t\rightarrow \lambda t, \qquad r\rightarrow \lambda r \qquad x_{i}\rightarrow \lambda x_{i}, \qquad y\rightarrow \lambda^{\frac{1}{z}} y, \qquad ds ^{2}\rightarrow \lambda^{\frac{2\theta}{dz}} ds ^{2}.
\end{equation}
The metric transforms covariantly under these transformations.

The holographic entanglement entropy of a curved region in this background is considered in \cite{Ghasemi:2019hdi}. The above metric has an anisotropic scaling in one of spatial coordinates, so we consider two possible ways to addressing this problem. First, we take the width of the strip along the anisotropic scaling direction $y$, while in the second case we put up the width of strip along one of the isotropic scaling directions $x_{i}$'s.

\subsubsection{The width of the strip along the anisotropic scaling direction}
In this case, we take the strip as
\begin{equation}
	\frac{-\ell}{2}\leq y \leq \frac{\ell}{2}, \qquad      0\leq x_{i} \leq H,         
\end{equation}
We choose the profile of the bulk minimal surface as $y=y(r)$. Hence, the induced metric on this minimal surface reads 
\begin{equation}
	ds^{2}=\tilde{L}^{2}r^{\frac{2\theta}{dz}-2}\Big(d\vec{x}_{d-1}^{2}+(1+\frac{y'^{2}}{r^{\frac{2}{z}-2}})dr^{2}\Big),
\end{equation}
where $\tilde{L}$ is the $AdS$ curvature scale, and $y'=\partial_{r}y$. The holographic entanglement entropy is given by

\begin{equation}\label{S-E}
	S_{EE}=\frac{2 \pi}{\ell_{p}^{d}}\int d\sigma\sqrt{\gamma}=\frac{\tilde{L}^{d}2 \pi H^{d-1}}{\ell_{p}^{d}}\int_{\delta}^{r_{m}}dr\frac{1}{r^{d+\frac{1-\theta}{z}-1}}\sqrt{r^{2(\frac{1}{z}-1)}+y'^{2}},
\end{equation}
in which $\delta$ is a $UV$ cut-off. $ r_ {m} $ is the turning point of the minimal surface defined in such a way that $r_{m}=r(x=0)$ and $y'(r_{m})=\infty$. Note that the entropy functional \eqref{S-E} does not depend explicitly on $y$, so there is a conserved quantity
\begin{equation}
	\frac{1}{r^{d+\frac{1-\theta}{z}-1}}\frac{y'}{\sqrt{r^{\frac{2}{z}-2}+y'^{2}}}=\frac{1}{r_{m}^{d+\frac{1-\theta}{z}-1}}
\end{equation}
Using this, we can rewrite $S_{EE}$ and width of the strip $\ell$ as functions of turning point $r_{m}$,
\begin{equation} 
	S_{EE}=\frac{\tilde{L}^{d}2 \pi H^{d-1}}{\ell_{p}^{d}}I,
\end{equation}
in which $I$ is
\begin{equation}\label{entropy-functional}
	I=\frac{1}{r_{m}^{d-\frac{\theta}{z}-1}}\int_{\frac{\delta}{r_{m}}}^{1}du\frac{u^{\frac{\theta}{z}-d}}{\sqrt{1-u^{2(d+\frac{1-\theta}{z}-1)}}},
\end{equation}
\begin{equation} 
	\ell=2r^{\frac{1}{z}}_{m}\int_{0}^{1}du\frac{u^{d+\frac{2-\theta}{z}-2}}{\sqrt{1-u^{2(d+\frac{1-\theta}{z}-1)}}},
\end{equation}
where $u=r/r_{m}$. If we explicitly compute the above integrals, we can write the $S_{EE}$ as a function of width of the strip $\ell$. To proceed, we must note that for special value of $\theta$ the logarithmic term appears. So, we separate it.

\textbf{i)}
$\theta\neq z(d-1)$
\begin{align}
	\ell&=\alpha r_{m}^{\frac{1}{z}}, \qquad  \alpha=2\sqrt{\pi}z\frac{\Gamma(\frac{z(d-1)+2-\theta}{2(z(d-1)+1-\theta)})}{\Gamma(\frac{1}{(2(z(d-1)+1-\theta))})} 
\end{align}

\begin{align}
	S_{EE}&=\frac{4\pi zH^{\frac{\theta}{z}}}{z(d-1)-\theta}\frac{\tilde{L}^{d}}{\ell_{p}^{d}}\Big(\frac{H}{\delta}\Big)^{d-\frac{\theta}{z}-1}-\beta_{d}\frac{\tilde{L}^{d}}{\ell_{p}^{d}}\Big(\frac{H}{\ell}\Big)^{z(d-1)-\theta}  \nonumber\\
\end{align}
where
\begin{align}
	\beta_{d}&=\frac{2z\pi^{\frac{3}{2}}}{2z(d-1)-2\theta+1}H^{(1-z)(d-1)+\theta}\Big(\frac{\Gamma(-\frac{1}{2}+\frac{1}{2(z(d-1)+1-\theta)})}{\Gamma(-1+\frac{1}{(2(z(d-1)+1-\theta))})} \Big)\alpha^{(d-1)z-\theta}.
\end{align}

\textbf{ii)}
$\theta= z(d-1)$

In this case, we find $\ell=2zr_{m}^{\frac{1}{z}}$, and the holographic entanglement entropy is 
\begin{align}
	S_{EE}&=2\pi H^{d-1}\frac{\tilde{L}^{d}}{\ell_{p}^{d}}\log  \frac{\ell^{z}}{(2z)^{z} \delta}.
\end{align}
This result shows that the boundary theory exhibits a logarithmic area law violation, which may be a sign that the boundary theory has a Fermi surface.\\

\subsubsection{The width of the strip along one of the isotropic scaling direction}
In this case, we take the strip as
\begin{equation}\label{metric2}
	\frac{-\ell}{2}\leq x_{1} \leq \frac{\ell}{2}, \qquad      0\leq x_{i} \leq H,\quad i\neq1 ,  \qquad      0\leq y \leq H,        
\end{equation}
We choose the profile of the bulk minimal surface as $x_{1}=x(r)$. So the induced metric on the minimal surface is derived as  
\begin{equation}
	ds^{2}=\tilde{L}^{d}r^{\frac{2\theta}{dz}-2}\Big(d\vec{x}_{d-2}^{2}+\frac{1}{r^{\frac{2}{z}-2}}dy^{2}+(1+x'^{2})dr^{2}\Big),
\end{equation}
where $\tilde{L}$ is the $AdS$ curvature scale, and $x'=\partial_{r}x$. The holographic entanglement entropy is given by

\begin{equation}\label{S-E2}
	S_{EE}=\frac{2 \pi }{\ell_{p}^{d}}\int d\sigma\sqrt{\gamma}=\frac{\tilde{L}^{d}2 \pi H^{d-1}}{\ell_{p}^{d}}\int_{\delta}^{r_{m}}dr\frac{\sqrt{1+x'^{2}}}{r^{d+\frac{1-\theta}{z}-1}},
\end{equation}
in which $\delta$ is a $UV$ cut-off, and $r_{m}$ is turning point of minimal surface, and $x'(r_{m})=\infty$. The corresponding conserved quantity is defined as
\begin{equation}
	\frac{1}{r^{d+\frac{1-\theta}{z}-1}}\frac{x'}{\sqrt{1+x'^{2}}}=\frac{1}{r_{m}^{d+\frac{1-\theta}{z}-1}}
\end{equation}
Using this we can rewrite the $S_{EE}$ and width of the strip $\ell$ as functions of turning point $r_{m}$,
\begin{equation} 
	S_{EE}=\frac{\tilde{L}^{d}2 \pi H^{d-1}}{\ell_{p}^{d}}I,
\end{equation}
in which $I$ is
\begin{equation}\label{entropy-functional2}
	I=\frac{1}{r_{m}^{d+\frac{1-\theta}{z}-2}}\int_{\frac{\delta}{r_{m}}}^{1}du\frac{1}{u^{d+\frac{1-\theta}{z}-1}}\frac{1}{\sqrt{1-u^{2(d+\frac{1-\theta}{z}-1)}}},
\end{equation}
\begin{equation} 
	\ell=2r_{m}\int_{0}^{1}du\frac{u^{d+\frac{1-\theta}{z}-1}}{\sqrt{1-u^{2(d+\frac{1-\theta}{z}-1)}}},
\end{equation}
Here $u=r/r_{m}$. Similarly, we can explicitly compute the above integrals, and rewrite the $S_{EE}$ as a function of width of the strip $\ell$. Again, for a special value of $\theta$ a logarithmic term appears. So, we separate it.

\textbf{i)}
$\theta\neq z(d-2)+1$
\begin{align}
	\ell&=\alpha r_{m}, \qquad  \alpha=2\sqrt{\pi}\frac{\Gamma(\frac{zd+1-\theta}{2(z(d-1)+1-\theta)})}{\Gamma(\frac{z}{(2(z(d-1)+1-\theta))})} 
\end{align}
\begin{align}
	S_{EE}&=\frac{4\pi z}{z(d-2)+1-\theta}\frac{\tilde{L}^{d}}{\ell_{p}^{d}}\Big(\frac{H}{\delta}\Big)^{d+\frac{1-\theta}{z}-2}-\beta_{d}\frac{\tilde{L}^{d}}{\ell_{p}^{d}}\Big(\frac{H}{\ell}\Big)^{d+\frac{1-\theta}{z}-2}  \nonumber\\
\end{align}
where
\begin{align}
	\beta_{d}&=\frac{2z\pi^{\frac{3}{2}}}{z(d-2)+1-\theta}H^{1-\frac{1}{z}}\Big(\frac{\Gamma(\frac{(2-d)z+\theta-1}{2(z(d-1)+1-\theta)})}{\Gamma(\frac{z}{(2(z(d-1)+1-\theta))})} \Big)\alpha^{\frac{(d-2)z+1-\theta}{z}}.
\end{align}

\textbf{ii)}
$\theta= z(d-2)+1$

In this case, we find $\ell=2r_{m}$, and the holographic entanglement entropy takes 
\begin{align}
	S_{EE}&\sim\frac{\tilde{L}^{d}}{\ell_{p}^{d}}\log  \frac{\ell}{ \delta}.
\end{align}
This result shows that the entanglement entropy of the boundary theory exhibits a logarithmic violation area law, which may be a sign that the boundary theory has a Fermi surface.\\

\section{A c-function from the holographic entanglement entropy in an anisotropic field theory}\label{sec:c-function}
In this section, we probe the monotonicity of a candidate $c$ function. Let us start with the following generic metric which exhibits violation of the Lorentz invariance and broken rotational symmetry, but respects translational symmetry,
\begin{align}\label{metric}
ds^2=e^{2A(r)}(-dt^2+dr^2+dx_{d-1}^2+e^{2h(r)}dy^2)
\end{align}
In section \ref{sec:examples}, we give an explicit example for such theories. To introduce the $c$-function, we consider a strip as the entangled region. Similar to the previous section, we study two cases where the strip is along one of  $x_i$'s and along $y$ direction in the following two subsections.

\subsection{The width of strip along one of the isotropic scaling dimensions} 
Now we need to derive the variation of $S$ with respect to $\ell$. We follow  \cite{Myers:2012ed} and details are left to the appendix \ref{app1}. It results to
\begin{align}\label{dS-dell3}
\frac{dS_x}{d\ell}&=\frac{2\pi H^{d-1}}{\ell_p^{d}}\frac{1}{K_d}.
\end{align}
Then define 
\begin{align}
c_x&:=\beta_x \frac{\ell^{d}}{H^{d-1}}\frac{\partial S_x}{\partial\ell}\nonumber\\
&=\beta_x \frac{2\pi\ell^{d}}{\ell_p^{d}K_d}
\end{align}
where $\beta_x$ is a positive numerical constant. We then find the flow $dc_x/dr_m$ where $r_m$ is defined as $r_m=r(0,\ell)$ with $\dot{r}(0,\ell)=0$ as
\begin{align}\label{RG-flow}
\frac{dc_x}{dr_m}&=\beta_x \frac{4\pi \ell^{d-1}}{\ell_p^{d}}B'(r_m)\int_0^{\ell/2} dx \frac{B'^2-d B''}{B'^2}
\end{align}
The details are given in the appendix \ref{app1}. 
Note that by setting $h(r)=0$ and changing the coordinate $r$ to domain wall coordinate, \eqref{RG-flow} transforms to the similar expression given in \cite{Myers:2012ed}.

To study the behaviour of \eqref{RG-flow} under the RG-flow, we use the null energy conditions (NECs). First, let us take a general null vector as 
\begin{align}
\xi=e^{-A}\Big[a_0\partial_t+a_r\partial_r+\sum_{i}^{d-1}a_i\partial_i+e^{-h}a_y\partial_y\Big]
\end{align}
with $a_0^2=a_r^2+a_y^2+\sum_{i=1}^{d-1}a_i^2$. Then, the general NEC can be read as
\begin{align}\label{general-NEC}
da_r^2A'^2-da_y^2A'h'-(a_r^2+a_y^2)h'^2-da_r^2A''-(a_r^2+a_y^2)h''\geq 0
\end{align}
One can put $a_r=1$ and $a_y=0$ to obtain 
\begin{align}\label{NEC1}
N_1\equiv dA'^2-h'^2-dA''-h'' \geq 0 
\end{align}
Another independent null energy condition can be found by taking $a_r=0$ and $a_y=1$,
\begin{align}\label{NEC2}
N_2 \equiv -d A' h'-h'^2-h'' \geq 0 
\end{align}
The inequalities \eqref{NEC1} and \eqref{NEC2} can be respectively transformed to
\begin{align}\label{NEC12}
	f_1'(r) &\equiv \Big(-B'e^{-dA+h}\Big)' = \Big(d(d-1)A'^2 + N_1\Big)e^{-dA+h} \geq 0, \\ \label{NEC22}
	f_2'(r) &\equiv \Big(-h'e^{B}\Big)' \geq 0. 
\end{align}
Now let us write the integrand of \eqref{RG-flow} as
\begin{align}\label{suff-cond-x}
	B'^2-d B''= d\,N_1 +(d-1) h'^2+2h'B'.
\end{align}
Since the first two terms in the right hand side are positive, it is enough to put $h'B'\geq 0$ to have sufficient condition for the integrand to be non-negative. Now \eqref{NEC12} and \eqref{NEC22} indicate that if one starts with positive $f_1(r)$ and $f_2(r)$ at the UV boundary, they remain positive in the bulk which means that we have $B'(r)\leq 0$ and $h'(r)\leq 0$ everywhere. Then from \eqref{suff-cond-x} the integrand of \eqref{RG-flow} is positive while $B'(r_m)\leq 0$ which in turn implies that the $c$-function is a monotonically decreasing function of $r_m$,
\begin{align}\label{dcdrm}
	\frac{dc_d}{dr_m} \leq 0.
\end{align} 
So using the NECs, the sufficient conditions are $B'(r)\leq 0$ and $h'(r)\leq 0$ at the UV boundary.


\begin{figure}
		\captionsetup{width=0.8\textwidth}
		\begin{center}
			\includegraphics[height=60mm]{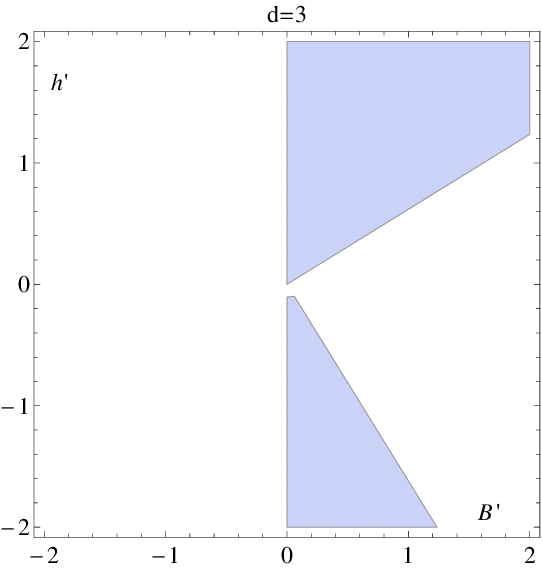} \\
			\caption{The shaded area in $B'-h'$ plane in $d=3$ show the regions where both $B'\geq 0$ and \eqref{NEC2} are satisfied. We also assumed $B''\geq 0$. }
			\label{hpBp}	
		\end{center}
\end{figure}

To satisfy \eqref{dcdrm}, another possibility is demanding $B'\geq 0$ and $B'^2-dB''\leq 0$. To have positive $B'$, it is sufficient to have  $f_1\leq 0$ at the IR limit. We also assume $B''\geq 0$ which implies $B'$ remains positive. Then 
\begin{align}
	B'^2-dB''= -N_1-(d+1)B''+2 B'^2-2 B' h'-(d-1) h'^2 
\end{align}
Since $N_1$ and $B''$ are positive, the sufficient condition is 
\begin{align}
2 B'^2-2 B' h'-(d-1) h'^2 \leq 0
\end{align}
Reminding our assumption $B'\geq 0$, the above inequality implies either $h'\geq B'(\sqrt{(2 d-1) }-1)/(d-1)  $ or $h'\leq B'(-\sqrt{(2 d-1) }-1)/(d-1)$. The corresponding region in the $B'-h'$ plane is shown in Fig. \ref{hpBp}. Unfortunately, neither IR nor UV boundary conditions are sufficients to satisfy these inequalities.

\subsection{The width of strip along the anisotropic direction}\label{subsec:y}
In this subsection, we consider the width of strip to be along $y$-direction (i.e., $-\ell/2 \leq y\leq \ell/2$). It is convenient to change variable to $d\rho=e^{-h}dr$, then the metric read as
\begin{align}\label{metric-y}
ds^2=e^{2A(\rho)+2h(\rho)}(-e^{-2h(\rho)}dt^2+d\rho^2+e^{-2h(\rho)}dx_{d-1}^2+dy^2)
\end{align}
By taking $\rho=\rho(y)$ as the Ryu-Takayanagi (RT) surface, the entropy functional would be  
\begin{align}\label{EE-y}
S_y=\frac{4\pi H^{d-1}}{\ell_p^{d}}\int_0^{\frac{\ell-\epsilon}{2}}dy e^{dA+h}\sqrt{\dot{\rho}^2+1}
\end{align} 
where $\dot{\rho}=d\rho/dy$. This is similar to \eqref{EE} by replacing $r \rightarrow \rho$ and everything should go exactly the same as previous subsection to find,
\begin{align}\label{RG-flow-y}
\frac{dc_y}{d\rho_m}&=\tilde{\beta}_y \frac{4\pi \ell^{d-1}}{\ell_p^{d}}B'(\rho_m)\int_0^\frac{\ell}{2} dy \frac{B'(\rho)^2-d B''(\rho)}{B'(\rho)^2}
\end{align}
Now we go back to $r$ coordinate,
\begin{align}\label{RG-flow-y-r}
	\frac{dc_y}{dr_m}&=\tilde{\beta}_y \frac{4\pi \ell^{d-1}}{\ell_p^{d}}B'(r_m)\int_0^\frac{\ell}{2} dy \frac{B'^2-d(h'B'+ B'')}{B'^2}
\end{align}
where all functions are in $r$ coordinate. 

Similar to previous subsection, we first consider $B'\leq 0$ while the integrand in \eqref{RG-flow-y-r} is positive. Then similar to \eqref{NEC1}, we can write
\begin{align}\label{suff-cond-y}
	B'^2-d (B''+h'B')= d\,N_1 +(d-1) h'^2+(2-d)h'B'
\end{align}
For $d=2$, this is always positive. For $d>2$ and since we assumed $B'\leq 0$, the sufficient condition is $h'\geq 0$. Eq. \eqref{NEC12} indicates that if starting with a negative $B'$ at UV boundary, it remains negative in whole regions inside the bulk. In contrast, there is no guarantee for $h'$ to remain positive starting with $h'>0$ at $r=0$. However, we show that our example in subsection \ref{sec:c-axion} satisfies conditions $B'<0$, $h'>0$ and gives monotically decreasing $c$-function in some range of parameters.

\begin{figure}
\captionsetup{width=0.8\textwidth}
	\begin{center}
		\includegraphics[height=55mm]{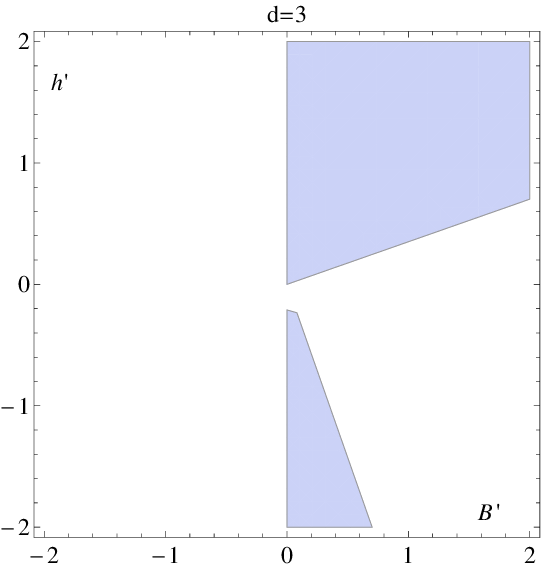}\\
		\caption{This diagram is related to the case that the width of strip is along $y$-direction with $d=3$. The shaded area in $B'-h'$ plane, shows the regions where both $B'\geq 0$ and \eqref{By} are valid. We also assumed $B''\geq 0$. }
		\label{fig:hpBp-y}	
	\end{center}
\end{figure}

Now consider the case, when $B'(r_m)\geq 0$ and the integrand of \eqref{RG-flow-y} is negative. We can write
\begin{align}
B'^2-d(B''+B'h')= -d N_1-2dB''+2 B'^2-B' (d+2) h'-(d-1) h'^2
\end{align} 
Again we assume $B''\geq 0$ and demand the right hand side of the above equation to be negative. It is sufficient to set
\begin{align}\label{By}
	2 B'^2-B' (d+2) h'-(d-1) h'^2\leq 0
\end{align}
The corresponding region in the  $B'-h'$ plane is shown in Fig. \ref{fig:hpBp-y}.\\

\section{Some examples}\label{sec:examples}
To ilustrate our RG-flow in \eqref{RG-flow}, we consider the axion-dilaton and hyperscaling models discussed in section \ref{sec:EE}. 

\subsection{c-function for the axion-dilaton model}\label{sec:c-axion}
\begin{figure}
\captionsetup{width=0.8\textwidth}
		\begin{center}
			\includegraphics[height=45mm]{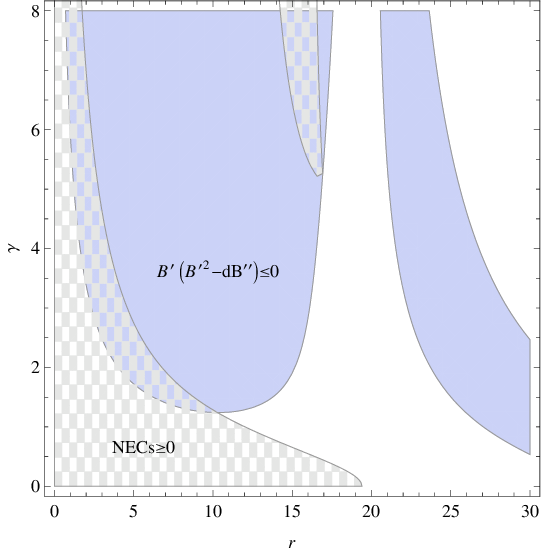}\,
			\includegraphics[height=45mm]{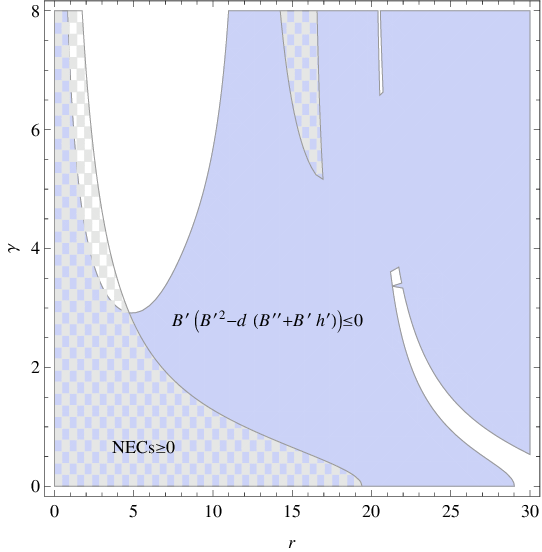}  \\
			\caption{In the left panel, shaded areas in $r-\gamma$ plane depict the regions where $B'^2-d B''$ and $B'$ have opposite signs. This is sufficient to have $dc_x/dr_m\leq 0$ for the width of strip along the $x$ direction. The shaded area in the right panel shows $dc_y/dr_m\leq 0$ for width of strip along the $y$ direction. The checkered regions in both panels show the area where $NEC's$ are valid. It can be infered that $dc_x/dr_m$ has no monotonic behavior, whereas, $dc_y/dr_m$ for $\gamma<3$ is monotonically decreasing as long as $NEC$'s are valid.}
			\label{small-ax}	
		\end{center}
\end{figure}

\begin{figure}
	\captionsetup{width=0.8\textwidth}
		\begin{center}
			\includegraphics[height=45mm]{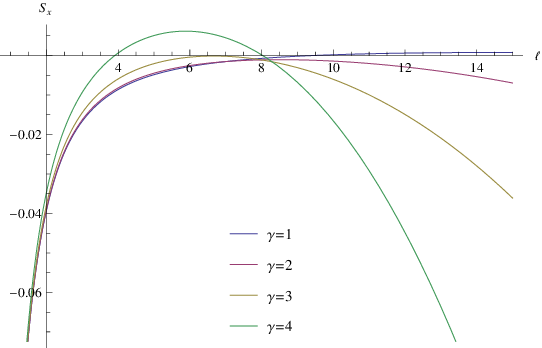} \;\;
			\includegraphics[height=45mm]{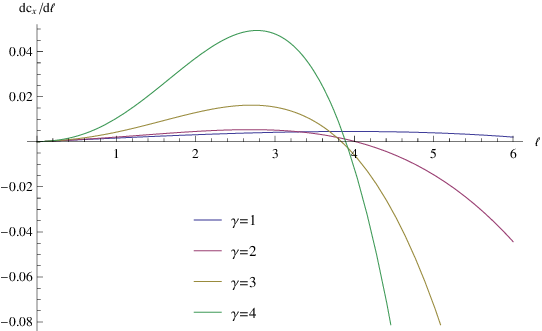} \\ \vspace{5mm}
			\includegraphics[height=45mm]{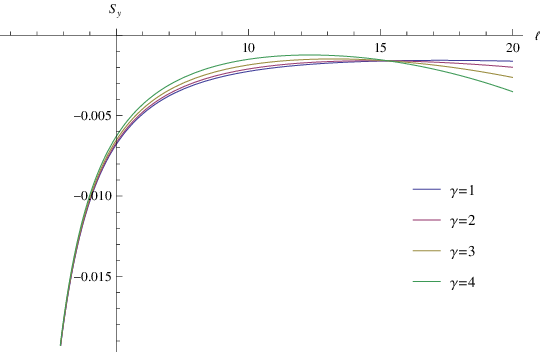} \;\;
			\includegraphics[height=45mm]{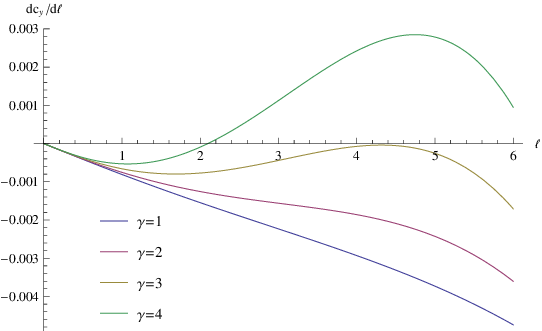} \\
			\caption{The above (below) panels are the entropy $S$ and the derivative of $c$-function for the width of strip in the $x$-direction ($y$-direction) as derived in appendix \ref{app:EE}. All diagrams are drawn for $a=0.1$ and $\gamma=1,2,3,4$. For the $x$ direction, there is no monotonically decreasing of the $c_x$-function, while in the $y$ direction we can see the monotonic behavior with negative derivative for $\gamma\leq 3$.}
			\label{S-dcdl}	
		\end{center}
\end{figure}

Let us consider the small $a$ expansion in \eqref{small-a} to find 
\begin{align}
	B(r)=-3 \log (r)+\frac{a^2 r^2}{12}-\frac{a^4r^4 }{64800}\left(1539 \gamma ^2+613\right)+\frac{a^4r^4}{120} \left(3 \gamma ^2+1\right)  \log (a r)+\cdots 
\end{align}
It indicates that near UV boundary, $B'(r)<0$, while from \eqref{small-a} we have $h'(r)>0$, so the sufficient conditions discussed below \eqref{suff-cond-x} are not satisfied. However, we can investigate the behavior of $dc/dr_m$ explicitly. For the width of strip along the $x$ direction, the integrand of \eqref{RG-flow} can be found as
\begin{align}
	\frac{B'^2-d B''}{B'^2}=-\frac{a^2 r^2}{6}-\frac{a^4 r^4}{972}  \left(54 \left(3 \gamma ^2+1\right) \log (ar)-81 \gamma ^2-22\right) +\cdots
\end{align}  
which is negative near UV boundary while $B'(r)<0$, therefore we have $dc_x/dr_m>0$ which has a wrong sign. In Fig. \ref{small-ax} left panel, the shaded area shows the region where $B'^2-d B''$ and $B'$ have opposite signs. It indicates that $dc_x/dr_m$ doesn't have any monotonic decreasing behavior. 
From the calculations in appendix \ref{app:EE}, we found $S_{EE}$ and $dc/d\ell$ as a function of $\ell$ and draw them in Fig. \ref{S-dcdl}. It shows that $dc/d\ell$ can not be negative in the UV regime.

For the width of strip along $y$-direction, the sufficient condition from \eqref{suff-cond-y} is $B'\leq 0$ and $h'\geq 0$. These are valid close to the UV boundary. To be sure about the behavior inside the bulk, let us compute the integrand of \eqref{RG-flow-y-r} as,
\begin{align}
	\frac{B'^2-d (B''+h'B')}{B'^2}=\frac{a^2 r^2}{12}+\frac{a^4 r^4}{3888} \left(37+81 \gamma ^2+108 \left(3 \gamma ^2+1\right) \log (ar)\right)
	 +\cdots
\end{align}
It is positive close to $r=0$ and since $B'<0$ in this region one expects $dc_y/dr_m$ to be negative. However, the right panel in Fig. \ref{small-ax} shows that this is valid in the region where $NEC$'s are valid and $\gamma\le 3$. For $\gamma>3$, there is a narrow region where $NEC$'s are valid, but the integrand of \eqref{RG-flow-y-r} is negative, which leads to the wrong sign $dc_y/dr_m>0$. One may claim that the integration in \eqref{RG-flow-y-r} from 0 to $r_m$ may gives an overall positive result, but the explicit investigation shows that this is not the case.

To summarize the results in this subsection, we found that the behaviour of $dc_x/dr_m$ does not decrease monotonically when the strip width is along the $x$-direction. In the case of $y$-direction, $dc_y/dr_m<0$ as long as $NEC$'s are valid and $\gamma\leq 3$. Explicit calculation of $c$-functions and its derivatives are given in appendix \ref{app:EE} and depicted in Fig. \ref{S-dcdl}.

\subsection{c-function for the hyperscaling model}\label{sec:c-hyperscaling}

Here, we consider the anisotropic model \eqref{axion-dilaton} with the potential \eqref{exp-potential}, which as discussed in the begining of subsection \ref{subsec:hyper}, is valid in the full range of scales from UV to IR. However, it can also be considered as the IR limit of \eqref{Full-potential}. This later point of view is more convenient to study the $c$-function, since our results in subsection \ref{subsec:hyper} indicate that the entanglement entropy behaves as $S\sim \ell^{1-n}$, where $n\neq d$ and depends on $d$, $\theta$ and $z$, so the $c$-function in \eqref{cd} does not approach to the CFT central chrage in the UV limit. As such, we consider \eqref{exp-potential} as the IR tail of the anisotropic theory and investigate the conditions for which the $c$-function \eqref{cd} is monotonincally decreasing along the RG-flow in this regime. 

Let us recall the solution for arbitrary $d$ as \cite{Giataganas:2017koz},
\begin{equation}\label{metric1-1}
	ds ^{2}= \tilde{L}^{2}r^{\frac{2\theta}{dz}}(\frac{-dt^{2}+dr^{2}+d\vec{x}_{d-1}^{2}}{r^{2}}+\frac{dy^{2}}{r^{\frac{2}{z}}}),
	\end{equation}
In comparison to \eqref{metric}, we identify metric functions as 
\begin{align}
A(r)=\Big(\frac{\theta}{dz}-1\Big)\log(r)\\
h(r)=\Big(1-\frac{1}{z}\Big)\log(r)
\end{align}
It follows then,
\begin{align}\label{Bp}
B'&=\frac{-d z+\theta +z-1}{z}\frac{1}{r} \\
\label{integrand}
I_x:=\frac{B'^2-dB''}{B'^2}&=\frac{d z}{-d z+\theta +z-1}+1
\end{align}
It is clear that both expressions have definite signs independent of $r$. So we can set parameters such that either $B'$ and $I_x $ have opposite signs and they never change their signs as functions of $r$. In this regime, the Null Energy Conditions (NEC) are as follows
\begin{align}
-d (\theta -1) z-d+\theta ^2 \geq 0 \\
(z-1) (d z-\theta +1) \geq 0
\end{align}

\begin{figure}
	\captionsetup{width=0.8\textwidth}
\begin{center}
	\includegraphics[height=55mm]{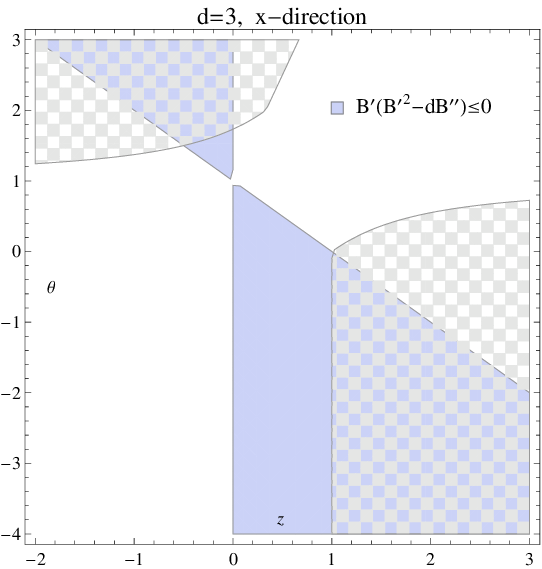} \;\;
	\includegraphics[height=55mm]{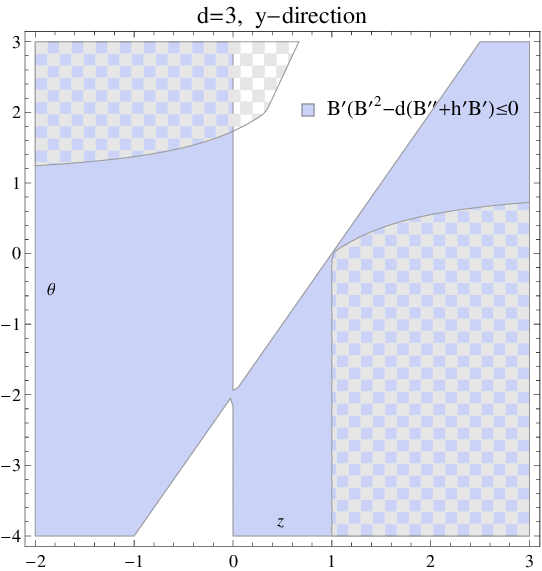} \\
	\caption{The left (right) panel is for the width of in the $x$-direction ($y$-direction) and the shaded areas in $\theta-z$ plane correspond to regions where  $B'$ and Eq. \eqref{integrand} (Eq. \eqref{integrandy}) have oposite signs hence $dc/dr_m\leq 0$. The checkered areas show the regions where $NEC$'s are valid. We set $d=3$.}
	\label{theta-z}	
\end{center}
\end{figure}

Fig. \ref{theta-z} left panel, shows regions in $\theta-z$ plane where $B'$ and $I_x $ have opposite signs and NEC's are satisfied which guarantees that $c_x$ is a  monotonically decreasing function of $r_m$. For the width of strip along the $y$-direction, we find the integrand of \eqref{RG-flow-y-r} as
\begin{align}\label{integrandy}
	I_y:=\frac{B'^2-d(B''+h'B')}{B'^2}&=\frac{z}{-d z+\theta +z-1}+1
\end{align}
Now we look for regions where $B'$ and $I_y$ have opposite signs which are the shaded regions shown in Fig. \ref{theta-z} right panel. This diagram shows that for $z>0$ and $\theta<0$, the $NEC's\geq 0$ are a subset of the shaded region which indicates that at least for $z>0$ and $\theta<0$, the null energy conditions are sufficient to have a monotonically decreasing $c$-function. 


\section{Conclusion}\label{sec:conclusion}
Within gauge/gravity duality, we calculate the entanglement entropy (EE) of a  strip-shaped region in $d+1$-dimensional quantum field theory (QFT) with broken Lorentz invariance as well as rotational symmetry. These theories are representative for many physical phenomena. So, extracting various information from these theories may be helpful in understanding of some quantum systems. One of the ways to receive information is computing the  entanglement entropy, which is a measure of entanglement structure of the ground state of underlying theory.

 We have computed the entanglement entropy of anisotropic strongly gauge fields which are dual to two classes of geometries. The first class is an anisotropic dilaton-axion background and the second one is identified with dynamical and hyperscaling violation parameters $\{z,\theta\}$. Note, these geometries are defined in some intermediate regime. We have done analytical computations in order to derived the holographic entanglement entropy of a strip-shaped region. 

 The holographic entanglement entropy of strip-shaped region has been used to construct a candidate $c$-funtion which is interpolated between UV and IR fixed points. We identified sufficient conditions to provide  monotonically decreasing of $c$ along RG flows. These conditions are extracted from the null energy conditions of the bulk matter fields.
 
 We considered two cases where the width of strip is along one of the isotropic directions or along the anisotropic one. In these two cases different conditions are needed to provide the monotonocity of $c$ function. For a strip with width along the anisotropic direction, the null energy conditions provided with suitable UV boundary conditions are sufficient to have a monotonically decreasing $c$-function. 

 In order to find an intuition of these general discussion, we illustrated the $c$ function in some special cases in five dimension. We have chosen two cases: near UV, and near IR expansions. For each one we observed that $c$-function monotonically decreased if the null energy conditions are satisfied. Orientation of the strip is also important. For the width of strip along the anisotropic direction the $NEC$'s are enough to have a monotically decreasing $c$-function, while for the other directions it is not so. This behavior is not so surprising, since the monotonicity of $c$-function is expected from the  Lorentz symmetry and subadditivity of the entanglement entropy. For an anisotropic theory the Lorentz symmetry is violated and there is no gaurantee for monotonicity of the $c$-function. 

It is worth notice that for theories that undergo a confinement/deconfinement transition, the behavior of $S(\ell)$ has generically a swallowtail behavior \cite{Gursoy:2018ydr}, (although this behavior should be present as well in the models  \cite{Giataganas:2017koz} for large values of sigma, where one also expects a deconfinement transition). Some portions of the swallowtail have $dS/d\ell<0$, $ d^2S/d\ell^2>0$. However, one can use the c-function argument to discard these branches. Indeed, one can show that the physical branches have both $c>0$ and $ dc/d\ell<0$ even though $dc/d\ell$ is discontinuous at the point of the transition\footnote{We would like to thank Juan F. Pedraza for pointing this out to us.}.

It would be also interesting to consider theories for which, in addition to the Lorentz-violation and broken rotation, there is no translation symmetries.

\vspace{10mm}
{\large{\bf Acknowledgment}}\\
 Special thanks to Juan F. Pedraza, for motivating this problem, useful comments and discussions. Authors also thank him for his generosity in sharing his perturbative calculations in section \ref{subsec:axion-dilaton}. We would like to thank Dimitrios Giataganas for useful comments and discussions. MG would like to thank Sepideh Mohammadi for encouragement and valuable comments.

\appendix

\section{Computing the entanglement entropy}\label{app:EE}
Starting with the entanglement region along one of the $x$- coordinates, we insert the solution \eqref{small-a} in \eqref{SEE} and expand in powers of $a$. Let us change the variable $u=r/r_m$ and subtract and add the divergent part as follows
\begin{align}\label{SEE-small-a}
	S_x&=\frac{4\pi H^{2}}{\ell_p^3}r_m\int_{0}^{1}du \Big[\frac{e^{2 B(r_mu)-B(r_m)}}{\sqrt{e^{2(B(r_mu)-B(r_m))}-1}}-\Big(\frac{a^2}{12 r_m u}+\frac{1}{r_m^3 u^3}\Big)\Big]\nonumber   \\
	&+\frac{4\pi H^{2}}{\ell_p^3}r_m\int_{\delta/r_m}^{1}du \Big(\frac{a^2}{12 r_m u}+\frac{1}{r_m^3 u^3}
	\Big)
\end{align}
The first integral can be expanded to
\begin{align}
	\frac{4\pi H^{2}}{\ell_p^3} \Big[\frac{1}{r_m^2}S_1+a^2S_2+a^4 r_m^2 S_3+a^4 r_m^2 S_4 \log (ar_m)  \Big]
\end{align}
where $S_i$'s are defined as follows and can be found by numerical integration, 
\begin{align}
	S_1&=\int_0^1du\Big[\frac{1}{\sqrt{\frac{1}{u^6}-1} u^6}-\frac{1}{u^3}\Big]=0.2844 \nonumber\\
	S_2&=\int_0^1du \Big[\frac{2 u^4+u^2+1}{12 \sqrt{\frac{1}{u^6}-1} u^4 \left(u^4+u^2+1\right)}-\frac{1}{12 u}
	\Big] =0.0351\nonumber\\
	S_3&=\int_0^1du  \Big[\frac{\left(1-u^2\right) \left(326 u^8+1552 u^6+1940 u^4+939 u^2+388\right)}{64800 \sqrt{\frac{1}{u^6}-1} \left(u^2-1\right) \left(u^5+u^3+u\right)^2} \nonumber\\
	&+\frac{\gamma ^2 \left(1-u^2\right) \left(3078 u^8+6156 u^6+7695 u^4+4617 u^2+1539\right)}{64800 \sqrt{\frac{1}{u^6}-1} \left(u^2-1\right) \left(u^5+u^3+u\right)^2} \nonumber\\
	&+\frac{540 \left(2 u^{10}+2 u^8+2 u^6-u^4-u^2-1\right)(1+3 \gamma ^2) \log (u)}{64800 \sqrt{\frac{1}{u^6}-1} \left(u^2-1\right) \left(u^5+u^3+u\right)^2}
	\Big]=-0.0073-0.0294 \gamma ^2 \nonumber\\
	S_4&=\int_0^1du  \Big[\frac{\left(2 u^5+2 u^3+u\right)(1+3 \gamma ^2)}{120 \left(u^4+u^2+1\right) \sqrt{1-u^6}}
	\Big] = 0.0085 \left(1+3 \gamma ^2\right) 
\end{align}
and the second integral in \eqref{SEE-small-a} is found to be
\begin{align}\label{div-part}
	\frac{4\pi H^{2}}{\ell_p^3}r_m\int_{\delta/r_m}^{1}du \Big(\frac{a^2}{12 r_m u}+\frac{1}{r_m^3 u^3}
	\Big)=\frac{4\pi H^{2}}{\ell_p^3}\Big(\frac{1}{2\delta ^2}-\frac{1}{2r_m^2}-\frac{1}{12} a^2 \log \left(\frac{\delta }{r_m}\right)\Big)
\end{align}
Putting all results together
\begin{align}\label{Srm}
	S_x=\frac{4\pi H^{2}}{\ell_p^3} \Big[&\frac{1}{r_m^2}S_1+a^2S_2+a^4 r_m^2 S_3+a^4 r_m^2 S_4 \log (ar_m) -\frac{1}{2r_m^2}+\frac{1}{12} a^2 \log (r_m)\Big] \nonumber\\
	=\frac{4\pi H^{2}}{\ell_p^3} \Big[&-(0.0073+0.0294 \gamma ^2) a^4 r_m^2+0.0085 \left(1+3 \gamma ^2\right) a^4 r_m^2 \log (ar_m) \nonumber\\
	&+0.0351 a^2-\frac{0.2156}{r_m^2} 
	+ \frac{a^2}{12} \log \left(r_m\right)\Big]
\end{align}
where we dropped the $\delta$ dependent divergent parts which are independent of $r_m$ and $\ell$ so don't contribute to the $c$-function. Notice that the last term is a universal term as it appears in the logarithmic part of \eqref{div-part}.

Similarly, we can compute the width of strip $\ell$ as a function of $r_m$. Start with
\begin{align}
	\ell&=2\int_0^{\ell/2}dx =2 \int_0^{r_m}\frac{dr}{|\dot{r}|}=2 \int_0^{r_m} dr \frac{1}{\sqrt{e^{2(B(r)-B(r_m))}-1}},
\end{align}
it follows then
\begin{align}\label{L}
	\ell&=2 r_m \left(L_1+a^2  r_m^2L_2 + a^4  r_m^4 L_3+a^4  r_m^4 \log (a r_m)L_4\right) \nonumber\\
	&=2 r_m \left(0.4312+0.0158 a^2 r_m^2-(0.0020+0.0062 \gamma ^2) a^4 r_m^4+0.0027 \left(1+3 \gamma ^2\right) a^4 r_m^4 \log (a r_m)\right)
\end{align}
where $L_i$'s are as follows
\begin{align}
	L_1&= \int_0^1 du \Big(\frac{1}{\sqrt{\frac{1}{u^6}-1}}
	\Big) =0.4312  \nonumber\\
	L_2&= \int_0^1 du \Big(\frac{1}{12 \sqrt{\frac{1}{u^6}-1} \left(u^4+u^2+1\right)}
	\Big) = 0.0158 \nonumber\\
	L_3&= \int_0^1 du \Big(\frac{-163 u^8-1063 u^6+838 u^2+540 \left(u^8+u^6+u^4\right) \log (u)+388}{64800 \sqrt{\frac{1}{u^6}-1} \left(u^2-1\right) \left(u^4+u^2+1\right)^2} \nonumber\\
	&+\frac{\gamma ^2 \left(-19 u^4+20 u^4 \log (u)+19\right)}{800 \sqrt{\frac{1}{u^6}-1} \left(u^6-1\right) }
	\Big) \nonumber\\
	&= -0.0020-0.0062 \gamma ^2 \nonumber\\
	L_4&= \int_0^1 du \Big(\frac{ \left(1+u^2\right)\left(1+3 \gamma ^2\right)}{120 \sqrt{\frac{1}{u^6}-1} \left(u^4+u^2+1\right)}\Big)  = 0.0027 \left(1+3 \gamma ^2\right)
\end{align}
Now we solve \eqref{L} perturbatively to find  $r_m $ in terms of  $\ell $,
\begin{align}
	r_m&=1.1596 \ell-0.0571 a^2 \ell^3+(0.0162+0.0185 \gamma ^2-0.0175 \gamma ^4) a^4 \ell^5   \nonumber\\
	&-(0.0012+0.0070 \gamma ^2+0.0105 \gamma ^4) a^4 \ell^5 \log (a\ell)
\end{align}  
Plugging in \eqref{Srm}, we obtain,
\begin{align} \label{Sell}
	S_x=\frac{4\pi H^{2}}{\ell_p^3} \Big[&-\frac{0.1603}{\ell^2}+\frac{a^2}{12}\log \ell
	+0.0316 a^2-(0.0088 +0.0293 \gamma ^2+ 0.0048 \gamma ^4) a^4 \ell^2 \nonumber\\
	&+(0.0079+0.0127 \gamma ^2-0.0326 \gamma ^4) a^4 \ell^2 \log (a\ell) \Big] 
\end{align} 
Here we can find $c_x$ from \eqref{cd} up to some constant,
\begin{align}
	c_x\sim \ell^3 \frac{\partial S_x}{\partial \ell} 
	&\sim 0.3207+0.0833 a^2 \ell^2-(0.0098+0.0459 \gamma ^2+0.0423 \gamma ^4)a^4 \ell^4 \nonumber\\
	&+\left(0.0157+0.0253 \gamma ^2-0.0653 \gamma ^4\right) a^4\ell^4 \log (a\ell)
\end{align} 
and its derivative
\begin{align}\label{dcxdl}
	\frac{\partial c_x}{\partial \ell} &\sim 0.1667 a^2 \ell 
	-(0.0236+0.1584 \gamma ^2+0.2345 \gamma ^4)a^4\ell^3
	\nonumber\\
	&+\left(0.0628+0.1015 \gamma ^2-0.2611 \gamma ^4\right)a^4\ell^3 \log (a \ell)
\end{align}

Similar calculations can be done for the region along $y$-direction. 
\begin{align}\label{SEE-small-a-y}
	S_y&=\frac{4\pi H^{2}}{\ell_p^3}r_m\int_{0}^{1}du \Big[\frac{e^{2 B(r_mu)-B(r_m)}}{e^{-h(r_mu)}\sqrt{e^{2B(r_mu)-2B(r_m)}-1}}-\Big(\frac{1}{r_m^3 u^3}-\frac{a^2}{24 r_m u}\Big)\Big]\nonumber   \\
	&+\frac{4\pi H^{2}}{\ell_p^3}r_m\int_{\delta/r_m}^{1}du \Big(\frac{1}{r_m^3 u^3}-\frac{a^2}{24 r_m u}	\Big)
\end{align}
We write the first integral as
\begin{align}
	\frac{4\pi H^{2}}{\ell_p^3} r_m\Big[\frac{1}{r_m^3}\tilde{S}_1+\frac{a^2}{r_m}\tilde{S}_2+a^4 r_m \tilde{S}_3+a^4 r_m \tilde{S}_4 \log (ar_m)
	\Big]
\end{align}
with, 
\begin{align}
	\tilde{S}_1&=\int_0^1du\Big[\frac{1}{\sqrt{\frac{1}{u^6}-1} u^6}-\frac{1}{u^3}\Big]=0.2844 \nonumber\\
	\tilde{S}_2&=\int_0^1du \Big[\frac{u^4-u^2-1}{24 \sqrt{\frac{1}{u^6}-1} u^4 \left(u^4+u^2+1\right)}+\frac{1}{24 u}
	\Big] =0.0062 \nonumber\\
	\tilde{S}_3&=\int_0^1du  \Big[\frac{-1579 u^{10}-2479 u^8+873 u^6+4279 u^4-221 u^2-873}{259200 \sqrt{\frac{1}{u^6}-1} \left(u^2-1\right) \left(u^5+u^3+u\right)^2} \nonumber\\
	&-\frac{1080 \left(u^{10}+u^8+u^6-3 u^4-3 u^2-3\right) \log (u)}{259200 \sqrt{\frac{1}{u^6}-1} \left(u^2-1\right) \left(u^5+u^3+u\right)^2}\nonumber\\
	&-\frac{\gamma ^2 \left(13 u^6+10 \left(u^6-3\right) \log (u)-19 u^2+6\right)}{800 \sqrt{\frac{1}{u^6}-1} u^2 \left(u^6-1\right)}
	\Big]=0.0033+0.0089 \gamma ^2 \nonumber\\
	\tilde{S}_4&=\int_0^1du  \Big[\frac{-(u^5+u^3+3u)(1+3 \gamma ^2)}{240 \left(u^4+u^2+1\right) \sqrt{1-u^6}}
	\Big] = -0.0061 \left(1+3 \gamma ^2\right) 
\end{align}
The second integral in \eqref{SEE-small-a-y} is found to be
\begin{align}\label{div-party}
	\frac{4\pi H^{2}}{\ell_p^3}r_m\int_{\delta/r_m}^{1}du \Big(\frac{1}{r_m^3 u^3}-\frac{a^2}{24 r_m u}	\Big)=\frac{4\pi H^{2}}{\ell_p^3}\Big(\frac{1}{24} a^2 \log \left(\frac{\delta }{r_m}\right)+\frac{1}{2\delta ^2}-\frac{1}{2r_m^2}\Big)
\end{align}
Then one finds
\begin{align}\label{Srmy}
	S_y=\frac{4\pi H^{2}}{\ell_p^3} \Big[&\frac{1}{r_m^2}\tilde{S}_1+a^2\tilde{S}_2+a^4 r_m^2 \tilde{S}_3+a^4 r_m^2 \tilde{S}_4 \log (ar_m)
	-\frac{1}{2r_m^2}-\frac{1}{24} a^2 \log (r_m)\Big] \nonumber\\
	=\frac{4\pi H^{2}}{\ell_p^3}\Big[&(0.0033+0.0089 \gamma ^2) a^4 r_m^2-0.0061 \left(1+3 \gamma ^2\right)  a^4 r_m^2 \log (a r_m)+0.0062 a^2 \nonumber\\
	&-\frac{0.2156}{r_m^2}
	- \frac{a^2}{24} \log \left(r_m\right)\Big] 
\end{align}
where we again dropped the $\delta$ dependent divergent parts.

We then compute the width of strip $\ell$ as a function of $r_m$ as follows,
\begin{align}
	\ell&=2\int_0^{\ell/2}dy =2 \int_0^{r_m}\frac{dr}{|\dot{r}|}=2 \int_0^{r_m} dr \frac{1}{e^{-h}\sqrt{e^{2B(r)-2B(r_m)}-1}} \nonumber\\
	&=2 r_m \left(\tilde{L}_1+a^2  r_m^2\tilde{L}_2 + a^4  r_m^4 \tilde{L}_3+a^4  r_m^4 \log (a r_m)\tilde{L}_4\right) \nonumber\\
	\label{Ly}
	&=(0.0051+0.0077 \gamma ^2 ) a^4 r_m^5-(0.0063+0.0189  \gamma ^2 ) a^4 r_m^5 \log (a r_m)-0.0517 a^2 r_m^3+0.8624 r_m
\end{align}
where $\tilde{L}_i$'s are given as
\begin{align}
	\tilde{L}_1&= \int_0^1 du \Big(\frac{1}{\sqrt{\frac{1}{u^6}-1}}
	\Big) =0.4312  \nonumber\\
	\tilde{L}_2&= \int_0^1 du \frac{-3 u^6-3 u^4-3 u^2+2}{24 \sqrt{\frac{1}{u^6}-1} \left(u^4+u^2+1\right)} = -0.0259 \nonumber\\
	\tilde{L}_3&= \int_0^1 du \Big[\frac{1552 + 6052 u^2 - 5125 u^4 - 9377 u^6 - 8477 u^8 + 5125 u^{10} + 
		5125 u^{12} + 5125 u^{14}}{259200 \sqrt{\frac{1}{u^6}-1} \left(u^2-1\right) \left(u^4+u^2+1\right)^2} \nonumber\\
	&\qquad\qquad-\frac{1080 u^4 \left(5 u^{10}+5 u^8+5 u^6-7 u^4-7 u^2-7\right) \log (u)}{259200 \sqrt{\frac{1}{u^6}-1} \left(u^2-1\right) \left(u^4+u^2+1\right)^2}\nonumber\\
	&\qquad\qquad+\frac{\gamma ^2 \left(25 u^{10}-44 u^4+10 \left(7-5 u^6\right) u^4 \log (u)+19\right)}{800 \sqrt{\frac{1}{u^6}-1} \left(u^6-1\right)} \Big]\nonumber\\
	&=0.0026+ 0.0038 \gamma ^2 \nonumber\\
	\tilde{L}_4&= \int_0^1 du \frac{(-5 u^8-5 u^6-5 u^4+2 u^2+2)(1+3\gamma^2)}{240 \sqrt{\frac{1}{u^6}-1} \left(u^4+u^2+1\right)}  = -0.00314521 \left(1+3 \gamma ^2\right)
\end{align}
Now we solve \eqref{Ly} perturbatively to find  $r_m $ in terms of  $\ell $,
\begin{align}
	r_m=1.1596 \ell+0.0935 a^2 \ell^3+(0.0124-0.0119 \gamma ^2) a^4 \ell^5+(0.0153+0.0459 \gamma ^2) a^4 \ell^5 \log (a\ell)
\end{align}  
Plugging in \eqref{Srmy}, we obtain,
\begin{align} \label{Selly}
	S_y=\frac{4\pi H^{2}}{\ell_p^3} \Big[&-\frac{0.1603}{\ell^2}-\frac{a^2}{24} \log (\ell)	
	+0.0259 a^2+(0.0002+0.0050 \gamma^2) a^4 \ell^2   \nonumber\\
	&-(0.0039+0.0118 \gamma ^2) a^4 \ell^2 \log (a\ell)\Big] 
\end{align} 
Then the $c$-function and its derivative can be found as
\begin{align}
	c_y\sim \ell^3 \frac{\partial S_y}{\partial \ell} 	
	&\sim 0.3207-0.0417 a^2 \ell^2-(0.0034+0.0018 \gamma ^2)a^4 \ell^4 \nonumber\\
	&-\left(0.0079+0.0236 \gamma ^2\right) a^4\ell^4 \log (a\ell)
	\nonumber\\
	\frac{\partial c_y}{\partial \ell} &\sim -0.0833 a^2 \ell 
	-(0.0216+0.0306 \gamma ^2)a^4\ell^3
	-\left(0.0314+0.0943 \gamma ^2\right)a^4\ell^3 \log (a \ell)    \label{dcydl}
\end{align}

\section{The RG-flow equation}\label{app1}
In this appendix, we present the details of deriving the RG-flow given in \eqref{RG-flow}.
First, we find variation of $S$ with respect to $\ell$, reminding that it has an explicit $\ell$ dependence through the integral upper bound and an implicit one through $r(x,\ell)$. One finds,
\begin{align}\label{dS-dell1}
\frac{dS}{d\ell}=\frac{4\pi H^{d-1}}{\ell_p^{d}}\frac{e^{(d-1)A+h}}{\sqrt{\dot{r}^2+1}}\Big[\frac{1}{2}\Big(1-\frac{d\epsilon}{d\ell}\Big)(\dot{r}^2+1)+\dot{r}\frac{\partial r}{\partial \ell}\Big]\Bigg\lvert_{x=\frac{\ell-\epsilon}{2}},
\end{align}
where we have used \eqref{EoM} and integration by parts. 
Following \cite{Myers:2012ed}, we take a variation of $r(\frac{\ell-\epsilon}{2},\ell)=r_c$ with respect to $\ell$ to find
\begin{align}
\Big[\frac{\dot{r}(x,\ell)}{2}\Big(1-\frac{d\epsilon}{d\ell}\Big)+\frac{\partial r(x,\ell)}{\partial \ell}\Big]_{x=\frac{\ell-\epsilon}{2}}=0.
\end{align}
Plugging in \eqref{dS-dell1}, we find
\begin{align}\label{dS-dell2}
\frac{dS}{d\ell}&=-\frac{4\pi H^{d-1}}{\ell_p^{d}}\frac{1}{K_d(\ell)}\frac{1}{\dot{r}}\frac{\partial r}{\partial \ell}\Bigg\lvert_{x=\frac{\ell-\epsilon}{2}}.
\end{align}
In the UV regime, $r\rightarrow 0$, the background is expected to be $AdS$, so we have,
\begin{align}
A(r)&\sim -\log(r) \nonumber\\
h(r)&\sim 0
\end{align}
Then we solve \eqref{K} to get $\dot{r}=-\sqrt{K_d^2 r^{2-2 d}-1}$ and integrate to find $x$,
\begin{align}\label{x}
x-\frac{\ell}{2}=\frac{1}{d-1}K_d^{\frac{1}{d-1}}F(K_d r^{1-d})
\end{align}
where 
\begin{align} 
F(u)=\int \frac{u^{\frac{d}{1-d}}}{\sqrt{u^2-1}}du
\end{align}
It is clear that at the boundary $F(u\rightarrow \infty)=0$. Now following \cite{Myers:2012ed}, take a partial derivative of \eqref{x} with respect to $\ell$, and solve for $\partial r/\partial \ell$, 
\begin{align}
\frac{\partial r}{\partial \ell}&=\frac{1}{(d-1)} \frac{ K_d'(\ell )}{K_d}r+\frac{1}{2 (d-1)^2} K_d^{\frac{2}{d-1}} \sqrt{K_d^2 r^{2-2 d}-1} \left((d-1)^2-2 K_d^{\frac{d}{1-d}} K_d'(\ell ) F(u)\right)
\end{align}
then divide by $\dot{r}=-\sqrt{K_d^2 r^{2-2 d}-1}$ and take the limit $r\rightarrow 0$ to find,
\begin{align}
\frac{1}{\dot{u}}\frac{\partial u}{\partial \ell}\Bigg\lvert_{x=\ell}=-\frac{1}{2}
\end{align}
It follows then
\begin{align}\label{dS-dell3-app}
\frac{dS}{d\ell}&=\frac{2\pi H^{d-1}}{\ell_p^{d}}\frac{1}{K_d}.
\end{align}
which is the equation \eqref{dS-dell3}. 

Now our task is to find the flow $dc_d/dr_m$ where $r_m$ is defined as $r_m=r(0,\ell)$ with $\dot{r}(0,\ell)=0$. We proceed as
\begin{align}\label{dc-drm}
\frac{dc_d}{dr_m}=\beta_d \frac{2\pi d\ell^{d-1}}{\ell_p^{d}K_d}\Big(\frac{d\ell}{dr_m}+\frac{1}{d}B'(r_m)\ell\Big)
\end{align}
On the other hand, $\ell$ can be found as a function of $r_m$,
\begin{align}
\ell&=2\int_0^{\ell/2}dx =-2 \int_0^{r_m}\frac{dr}{\dot{r}}=2 \int_0^{r_m}dr \frac{1}{\sqrt{K_d^2e^{2B}-1}}
\end{align}
Now define $w=K_de^{B}$ and
\begin{align}
G(w):=\int \frac{w^{-(1+\frac{1}{d})}}{\sqrt{w^2-1}}dw
\end{align}
which is indeed a hypergeometric function with $G(w\rightarrow \infty)=0$ at the boundary $\rightarrow 0$. Then we have
\begin{align}\label{ell}
\ell&=2\int_{r=0}^{r=r_m} \frac{K_d^{1/d}e^{B/d}}{B'} dG \nonumber\\
&=2\frac{G(1)}{B'(r_m)} -2\int_0^{r_m}K_d^{1/d}G \Big(\frac{e^{B/d}}{B'}\Big)'dr
\end{align}
where the first term is the boundary term. 
We now take a derivative with respect to $r_m$,
\begin{align}\label{dell-drm}
\frac{d\ell}{dr_m}&=-\frac{2}{d}G(1) +
\frac{2}{d}B'(r_m)\int_0^{r_m}K_d^{1/d}G \Big(\frac{e^{B/d}}{B'}\Big)'dr 
- 2\int_0^{r_m} K_d^{1/d}\frac{dG}{dr_m} \Big(\frac{e^{B/d}}{B'}\Big)'dr \nonumber\\
&= -\frac{2}{d}G(1)+\frac{2}{d}B'(r_m)\int_0^{r_m} \Big(K_d^{1/d} G(w)+\frac{de^{-B/d}}{\sqrt{w^2-1}}\Big)
\Big(\frac{e^{B/d}}{B'}\Big)'dr
\end{align}
in which we have used
\begin{align}
\frac{dG}{dr_m}=-e^{B}K_dB'(r_m)\frac{dG}{dw}
\end{align}
Now plugging \eqref{ell} and \eqref{dell-drm} in \eqref{dc-drm} we obtain,
\begin{align}
\frac{dc_d}{dr_m}&=\beta_d \frac{4\pi d\ell^{d-1}}{\ell_p^{d}}B'(r_m)\int_0^{r_m} \frac{e^{-B/d}}{\sqrt{K_d^2e^{2B}-1}}\Big(\frac{e^{B/d}}{B'}\Big)'dr
\nonumber\\
&=\beta_d \frac{4\pi d\ell^{d-1}}{\ell_p^{d}}B'(r_m)\int_0^{\ell/2} dx e^{-B/d}\Big(\frac{e^{B/d}}{B'}\Big)'
\nonumber\\
&=\beta_d \frac{4\pi \ell^{d-1}}{\ell_p^{d}}B'(r_m)\int_0^{\ell/2} dx \frac{B'^2-d B''}{B'^2}
\end{align}


\end{document}